\documentclass[a4paper, 12pt]{article}
\usepackage[margin={30mm,45mm},verbose]{geometry}
\usepackage{graphicx}
\usepackage{amsmath}
\usepackage{amssymb}
\usepackage[colorlinks=true, urlcolor=blue]{hyperref}
\begin{document}
\title{Truncated Mellin moments: Useful relations and implications for the
spin structure function $g_2$ 
}
\author{Dorota Kotlorz\footnote{Opole University of Technology,
E-mail: {\tt d.strozik-kotlorz@po.opole.pl}} and Andrzej Kotlorz\\
Opole University of Technology}
\maketitle
\abstract{
We review our previous studies of truncated Mellin moments of parton
distributions. We show in detail the derivation of the evolution equation
for double truncated moments. The obtained splitting function has the same
rescaled form as in a case of the single truncated moments. We apply the
truncated moments formalism to QCD analyses of the spin structure functions
of the nucleon, $g_1$ and $g_2$. We generalize the Wandzura-Wilczek relation
in terms of the truncated moments and find new sum rules. We derive the
DGLAP-like evolution equation for the twist-2 part of $g_2$ and solve it
numerically. We find also useful relations between the truncated and
untruncated moments.}\\ \\
PACS {12.38.Bx 11.55.Hx}\\
Perturbative calculations, Sum rules, Truncated moments
  
\section{Introduction}
Truncated moments (TM) of parton distribution functions (PDFs) were
introduced and developed in the QCD analysis by S.~Forte, J.~Latorre,
L.~Magnea, A.~Piccione and G.~Ridolfi \cite{b1}-\cite{b4}.
The authors obtained the non-diagonal evolution equations, where
each $n$th truncated moment couples to all higher ones.  Then, the idea of
TM was successfully applied in the leading $ln^2x$ approximation, where we
found diagonal solutions \cite{b5}. Also A.~Sissakian, O.~Shevchenko and
O.~Ivanov used the TM technique in their NLO analyses of SIDIS data,
incorporating polynomial expansion \cite{b6}, \cite{b7}. Several years ago,
we derived the DGLAP-type diagonal (no mixing between moments of different
orders) and exact evolution equation for the TM in a case of a single
truncation \cite{b8}. Then, we have utilized this approach to the determination
of the parton distribution functions from their truncated moments \cite{b10}.
The idea of the TM was also discussed by A. Psaker, W. Melnitchouk,
M. E. Christy and C. Keppel in a context of the quark-hadron duality \cite{b9}.
In this paper, in the continuation of our earlier works, we present in detail
a generalization of the TM approach for a double truncation of
the integral limits. This problem has been already briefly discussed in
\cite{b9} and presented in \cite{b11}. Here, we also present interesting,
novel implications of the TM approach for the analysis of the polarized
structure function $g_2$.

The evolution equations for the truncated Mellin moments of the parton
distributions can be a useful additional tool in the perturbative QCD
analysis of the structure functions.
In the standard Dokshitzer-Gribov-Lipatov-Altarelli-Parisi (DGLAP) formalism
\cite{b12}-\cite{b15}, a central role play the parton densities, which depend
on the kinematic variables $Q^2$ and $x$. Then the truncated or untruncated
moments, which are e.g. contributions to the sum rules, can be obtained
by integrating of the parton distribution $q(x,Q^2)$ over the Bjorken-$x$.
Alternatively, one can study directly the $Q^2$ evolution of the moments.
This is sometimes more convenient, particularly in the cases when we know the
moments (e.g. from direct measurements, calculations on lattice and sum rules
constraints), while the PDFs are poorly known. 
We have shown in \cite{b8} that the evolution equation for the $n$th truncated
at $x_0$ moment $\int\limits_{x_0}^1 dx\, x^{n-1}\, q(x,Q^2)$ has the same
DGLAP form as for the parton density itself, but with a modified splitting
function $P_{ij}'(n,x)= x^n P_{ij}(x)$. 
The TM approach allows one to avoid the problem of the unphysical region
$x\rightarrow 0$. Furthermore, this approach refers directly to the physical
values - moments (rather than to the parton distributions), what enables one to
use a wide range of deep-inelastic scattering data in terms of a smaller number
of parameters. The evolution equations for the truncated moments are universal -
they can be used in each order of the approximation (LO, NLO, NNLO etc.) and
for unpolarized, as well as polarized parton densities.

The contents of this paper are as follows. In Section 2 we recall the
evolution equation for the $n$th truncated at $x_0$ moment and generalize
this formula for the double truncation - in both limits of integration.
Novel implications of the TM approach for the the polarized
structure function $g_2$ are presented in Section 3. We derive the
Wandzura-Wilczek relation in terms of the truncated moments, finding new sum
rules. Then we derive the DGLAP-like evolution equation for
the twist-2 part of $g_2$ and present numerical solutions.
In Section 4 we derive some useful relations between the truncated and
untruncated Mellin moments. Finally we summarize the main results and discuss
future possible applications of the TM approach.

\section{The evolution equations for the truncated Mellin moments
of the parton densities}

The standard perturbative QCD approach is based on the
Dokshitzer-Gribov-Lipatov-Altarelli-Parisi (DGLAP) evolution equations for
the parton densities \cite{b12}-\cite{b15}. In this formalism the main role
is played by the parton distribution functions $q(x,Q^2)$, which obey the
well-known formula
\begin{equation}\label{eq.2.1}
\frac{dq(x,Q^2)}{d\ln Q^2}=\frac{\alpha_s(Q^2)}{2\pi}\; (P\otimes q)(x,Q^2),
\end{equation}
where
$\alpha_s(Q^2)$ is the running coupling and $\otimes$ denotes the Mellin
convolution
\begin{equation}\label{eq.2.3}
(A\otimes B)(x)\equiv\int\limits_{x}^{1} \frac{dz}{z}\,
A\left(\frac{x}{z}\right)\,B(z).
\end{equation}
The splitting function $P(z,Q^2)$ can be expanded in a power series of
$\alpha_s(Q^2)$.\\
For the untruncated $n$th Mellin moment, defined for an arbitrary function
$f(x)$ as
\begin{equation}\label{eq.2.4}
\bar{f}^{n}=\int\limits_0^1 dx\, x^{n-1}\, f(x),
\end{equation}
the evolution equation takes the form of an ordinary linear differential
one, namely
\begin{equation}\label{eq.2.5}
\frac{d\bar{q}^{n}(Q^2)}{d\ln Q^2}=\frac{\alpha_s(Q^2)}{2\pi}\;
\gamma^n(Q^2)\, \bar{q}^{n}(Q^2).
\end{equation}
Here, the anomalous dimension $\gamma^n(Q^2)$ is the untruncated moment of
the splitting function $P(z,Q^2)$:
\begin{equation}\label{eq.2.6}
\gamma^n(Q^2)=\int\limits_0^1 dz\, z^{n-1}\, P(z,Q^2).
\end{equation}
Eq.~(\ref{eq.2.5}) can be solved analytically:
\begin{equation}\label{eq.2.6a}
\bar{q}^n(Q^2) = 
\bar{q}^n(Q_0^2)\left[\frac{\alpha_s(Q_0^2)}{\alpha_s(Q^2)}\right]^
{\,b\,\gamma^n}
\end{equation}
and the parton density $q(x,Q^2)$
can be found via the inverse Mellin transform
\begin{equation}\label{eq.2.7}
q(x,Q^2)=\frac{1}{2\pi i}\int\limits_{c-i\infty}^{c+i\infty} dn\,
x^{-n}\, \bar{q}^n(Q^2).
\end{equation}
Now let us focus on the approach, in which we can study the $Q^2$
evolution of the truncated moments of the parton distributions.
In \cite{b8} we have found that the single truncated moments of the
parton distributions $q(x,Q^2)$, defined as
\begin{equation}\label{eq.2.8}
\bar{q}^{n}(x_0,Q^2)=\int\limits_{x_0}^1 dx\, x^{n-1}\, q(x,Q^2),
\end{equation}
obey the DGLAP-like equation\footnote{For clarity we present only the
nonsinglet part}
\begin{equation}\label{eq.2.9}
\frac{d\bar{q}^n(x_0,Q^2)}{d\ln Q^2}=
\frac{\alpha_s(Q^2)}{2\pi}\; (P'\otimes \bar{q}^n)(x_0,Q^2).
\end{equation}
Here a role of the splitting function plays $P'(n,z)$:
\begin{equation}\label{eq.2.10}
P'(n,z)= z^n\, P(z).
\end{equation}
Since the experimental data cover only a limited range of $x$, except very
small $x\rightarrow 0$ as well as large $x\rightarrow 1$, it is very natural
and convenient to deal with the double truncated moments.
Truncation at large $x$ is less important in comparison to the small-$x$
limit because of the rapid decrease of the parton densities as
$x\rightarrow 1$, nevertheless a comprehensive theoretical analysis requires
an equal treatment of the both truncated limits.

It can be shown (see \cite{b9}, \cite{b11}), that the double truncated moments
\begin{equation}\label{eq.2.11}
\bar{q}^{n}(x_{min},x_{max},Q^2)=
\int\limits_{x_{min}}^{x_{max}} dx\, x^{n-1}\, q(x,Q^2)
\end{equation}
also satisfy the DGLAP-type evolution Eq.~(\ref{eq.2.9}), namely
\begin{equation}\label{eq.2.12}
\frac{d\bar{q}^n(x_{min},x_{max},Q^2)}{d\ln Q^2} =
\frac{\alpha_s(Q^2)}{2\pi}\;
\int\limits_{x_{min}}^{1}\frac{dz}{z}\; P'(n,z)\;
\bar{q}^n\left( \frac{x_{min}}{z}, \frac{x_{max}}{z}, Q^2 \right)
\end{equation}
with $P'$ given again by Eq.~(\ref{eq.2.10}).
Indeed, one can note that the double truncated moment, defined
by Eq.~(\ref{eq.2.11}), is a subtraction of two single truncated ones
Eq.~(\ref{eq.2.8}):
\begin{eqnarray}\label{eq.A.1}
\bar{q}^{n}(x_{min},x_{max},Q^2) &=&
\int\limits_{x_{min}}^{1} dx\, x^{n-1}\, q(x,Q^2) - 
\int\limits_{x_{max}}^{1} dx\, x^{n-1}\, q(x,Q^2) \nonumber\\
&=& \bar{q}^{n}(x_{min},Q^2) - \bar{q}^{n}(x_{max},Q^2).
\end{eqnarray} 
Applying the evolution equation for the single truncated moments
Eq.~(\ref{eq.2.9}) to the above formula, one can write
\begin{eqnarray}\label{eq.A.2}
\frac{d\bar{q}^n(x_{min},x_{max},Q^2)}{d\ln Q^2}=
\frac{\alpha_s(Q^2)}{2\pi}\;\Bigg[\:
\int\limits_{x_{min}}^{1}\frac{dz}{z}\; P'(n,z)\;
\bar{q}^n\left( \frac{x_{min}}{z}, Q^2 \right) \nonumber\\
- \int\limits_{x_{max}}^{1}\frac{dz}{z}\; P'(n,z)\;
\bar{q}^n\left( \frac{x_{max}}{z}, Q^2 \right)\:\Bigg].
\end{eqnarray}
Then the lower limit of integration in the second integral on the r.h.s. can
be moved from $x_{max}$ to $x_{min}$ since for $z$ below
$x_{max}$ the argument of the truncated moment becomes greater then one and
such a moment is equal to zero. This leads immediately to
Eq.~(\ref{eq.2.12}).   
Our approach Eq.~(\ref{eq.2.9}) - Eq.~(\ref{eq.2.12}) is valid for the coupled
DGLAP equations for quarks and gluons and for any approximation (LO, NLO,
NNLO, etc.). Only for clarity we present here the nonsinglet and leading
order part. Let us emphasize that the evolution equations for the double
truncated moments Eq.~(\ref{eq.2.12})
are in fact a valuable generalization of those for the single truncated and
untruncated ones. Setting $x_{min} = x_0$ or $x_{min} = 0$ and $x_{max} = 1$,
one obtains Eq.~(\ref{eq.2.9}) or Eq.~(\ref{eq.2.5}), respectively.
In the next section we study the application of the presented equations to
the polarized structure function $g_2$.

\section{Predictions for the spin structure function $g_2$, based on the
truncated moments}

For a complete description of the nucleon spin, one needs two polarized
structure functions: $g_1$ and $g_2$. Recently, a new generation of
experiments with high polarized luminosity, performed at Jefferson Lab,
allows more precise study of the polarized structure functions and their
moments. This is crucial in our understanding of the QCD spin sum rules,
higher-twist effects and quark-hadron duality. 

The function $g_1$ has a simple interpretation in the parton model:
\begin{equation}\label{eq.3.1}
g_1(x) = \frac{1}{2}\sum_i e_i\, \Delta q_i(x),
\end{equation}
describing the distribution of quark spin in the nucleon, while function $g_2$
has no such physical meaning in this classic model. Due to the technical
difficulties of obtaining transversely polarized targets, the structure function
$g_2$ has not been a topic of investigations for a long time. Recently, new
experimental data at low and intermediate momentum transfers make $g_2$ also
a valuable and hopeful tool to study the spin structure of the nucleon.
The function $g_2$ provides knowledge on higher twist effects, which are
reflection of the quark-gluon correlations in the nucleon. A particular
important role in this analysis is played by moments of the spin structure
functions
\begin{equation}\label{eq.3.2}
\Gamma_1^n = \int_0^1 dx\, x^{n-1} g_1(x,Q^2),
\end{equation}
\begin{equation}\label{eq.3.3}
\Gamma_2^n = \int_0^1 dx\, x^{n-1} g_2(x,Q^2).
\end{equation}
They are a sensitive tool for testing the QCD sum rules and determination of the
higher twist contributions (for a review of this problem see, e.g., \cite{b16}).
Here, we would like to focus on the application of the TM and their
evolution equations to predictions for the spin structure function $g_2$.

The experimental value of the function $g_2$, measured in the small to
intermediate $Q^2$ region, consists of two parts: the twist-2 (leading) and
the higher twist term:
\begin{equation}\label{eq.3.4}
g_2(x,Q^2) = g_2^{LT}(x,Q^2) + g_2^{HT}(x,Q^2).
\end{equation}
The leading-twist term $g_2^{LT}$ can be determined from the other structure
function - $g_1$ via the Wandzura-Wilczek relation \cite{b17}
\begin{equation}\label{eq.3.5}
g_2^{LT}(x,Q^2) = g_2^{WW}(x,Q^2) = -g_1(x,Q^2) + \int_x^1 \frac{dy}{y}\,
g_1(y,Q^2).
\end{equation}
Then, from the measurements of $g_1$ and $g_2$, using the Wandzura-Wilczek
approximation Eq.~(\ref{eq.3.5}), one is able to extract the higher-twist
term $g_2^{HT}$.
We find a new equation which is a generalization of the Wandzura-Wilczek relation,
for the truncated moments:
\begin{equation}\label{eq.3.7}
\bar{g}_2^n(x_0,Q^2) = \frac{1-n}{n}\: \bar{g}_1^n(x_0,Q^2)
- \frac{x_0^n}{n}\:\bar{g}_1^0(x_0,Q^2). 
\end{equation}
It is easy to see that for the untruncated moments Eq.~(\ref{eq.3.7}) takes
the well known form
\begin{equation}\label{eq.3.6}
\bar{g}_2^n(Q^2) = \frac{1-n}{n}\: \bar{g}_1^n(Q^2). 
\end{equation}
Here,
\begin{equation}\label{eq.3.8}
\bar{g}_{1,2}^{n}(Q^2)=\int\limits_{0}^1 dx\, x^{n-1}\, g_{1,2}(x,Q^2),
\end{equation}
\begin{equation}\label{eq.3.9}
\bar{g}_{1,2}^{n}(x_0,Q^2)=\int\limits_{x_0}^1 dx\, x^{n-1}\, g_{1,2}(x,Q^2),
\end{equation}
and consequently
\begin{equation}\label{eq.3.10}
\bar{g}_1^{0}(x_0,Q^2)=\int\limits_{x_0}^1 \frac{dx}{x}\, g_1(x,Q^2).
\end{equation}
For the first moment ($n=1$) Eq.~(\ref{eq.3.7}) reads
\begin{equation}\label{eq.3.11}
\bar{g}_2^1(x_0,Q^2) = - x_0\:\bar{g}_1^0(x_0,Q^2) 
\end{equation}
or equivalently
\begin{equation}\label{eq.3.12}
\int\limits_{x_0}^1 dx\, g_2(x,Q^2) =
-x_0\int\limits_{x_0}^1 \frac{dx}{x}\,g_1(x,Q^2).
\end{equation}
The moments of the spin structure functions $g_1$ and $g_2$ are of great
importance due to their relations to the fundamental QCD sum rules:
\begin{itemize} 
\item
Bjorken sum rule (BSR) \cite{b18}
\begin{equation}\label{eq.3.13}
\int\limits_0^1 dx\,\left[ g_1^p(x,Q^2) - g_1^n(x,Q^2) \right ] =
\frac{g_A}{6},
\end{equation}
\item
Efremov-Leader-Teryaev sum rule (ELT) \cite{b19}
\begin{equation}\label{eq.3.14}
\int\limits_0^1 dx\, x\,\left[ g_2^p(x,Q^2) - g_2^n(x,Q^2) \right ] =
-\frac{1}{2}\int\limits_0^1 dx\, x\,\left[ g_1^p(x,Q^2) - g_1^n(x,Q^2) \right ],
\end{equation}
\item
Burkhardt-Cottingham sum rule (BC) \cite{b20}
\begin{equation}\label{eq.3.15}
\int\limits_0^1 dx\, g_2(x,Q^2) = 0.
\end{equation}
\end{itemize}
From  Eq.~(\ref{eq.3.7}), setting $n=1$ and $x_0\rightarrow 0$, one can obtain
automatically the BC sum rule for $g_2^{WW}$. Furthermore, using the
generalization of the Wandzura-Wilczek relation Eq.~(\ref{eq.3.7})
for $n=1$ at two different points of the truncation and also applying the BC
sum rule Eq.~(\ref{eq.3.15}), we obtain an interesting relation:
\begin{equation}\label{eq.3.16}
\int\limits_{x_1}^{x_2} dx\, g_2^{WW}(x,Q^2) =
(x_2-x_1)\int\limits_{x_2}^1\frac{dx}{x}\,g_1(x,Q^2)-
x_1\int\limits_{x_1}^{x_2}\frac{dx}{x}\, g_1(x,Q^2).
\end{equation}
The above formula can be very useful in determination of the partial twist-2
contribution to the BC sum rule. For example, setting $x_1=0$ and $x_2=x_0$,
when $x_0\rightarrow 0$, one can get the small-$x$ contribution to the BC
sum rule:  
\begin{equation}\label{eq.3.17}
\int\limits_{0}^{x_0} dx\, g_2^{WW}(x,Q^2) =
x_0\int\limits_{x_0}^1\frac{dx}{x}\,g_1(x,Q^2).
\end{equation}
Since it is assumed that for $x<0.02$ the higher-twist effects are negligible
(see, e.g., \cite{b21}), Eq.~(\ref{eq.3.17}) allows one to estimate
the small-$x$ ($x_0<0.02$) contribution as
\begin{equation}\label{eq.3.18}
\int\limits_{0}^{x_0<0.02} dx\, g_2(x,Q^2) \approx
\int\limits_{0}^{x_0<0.02} dx\, g_2^{WW}(x,Q^2).
\end{equation}
Concluding, the obtained Eqs.~(\ref{eq.3.7}), (\ref{eq.3.12}),
(\ref{eq.3.16}) and (\ref{eq.3.17}) are some sorts of sum rules and can be
significant in increasing of the reliability of the data analyses in spin
physics. 

Now we would like to discuss the problem of the $Q^2$ evolution of $g_2$.
While a general DGLAP-type equation for $g_2$ does not exist, for the twist-3
component of $g_2$ suitable evolution equations have been formulated by
V.~M.~Braun, G.~P.~Korchemsky and A.~N.~Manashov in
\cite{b21}-\cite{b23}. In the leading twist-2 approximation, the $Q^2$
evolution of $g_2$ is governed by the evolution of $g_1$, according to the
Wandzura-Wilczek relation.
Since the second term on the r.h.s. of Eq.~(\ref{eq.3.5}) is the $n=0$th
truncated moment of the function $g_1$ Eq.~(\ref{eq.3.10}), we can rewrite
the Wandzura-Wilczek relation in the form
\begin{equation}\label{eq.3.21}
g_2^{WW}(x,Q^2) = -g_1(z,Q^2) + \bar{g}_1^0(z,Q^2)
\end{equation}
and obtain the evolution equation for $g_2^{WW}$:
\begin{equation}\label{eq.3.19}
\frac{dg_2^{WW}(x,Q^2)}{d\ln Q^2} = -\frac{dg_1(x,Q^2)}{d\ln Q^2}+
\frac{d\bar{g}_1^0(x,Q^2)}{d\ln Q^2}.
\end{equation}
It is worth noting that according to Eqs.~(\ref{eq.2.9}),(\ref{eq.2.10}),
the $n=0$th truncated moment of the parton distribution $q$ evolves in the
same way as $q$ itself ($P'(0,z)=P(z)$). Taking this into account in the
case of $g_1$, we obtain from Eqs.~(\ref{eq.3.19}), (\ref{eq.3.21}) the
evolution equation
\begin{equation}\label{eq.3.20}
\frac{dg_2^{WW}(x,Q^2)}{d\ln Q^2} = \frac{\alpha_s(Q^2)}{2\pi}
\int\limits_x^1 \frac{dz}{z}\,P\left(\frac{x}{z}\right)\left[
\bar{g}_1^0(z,Q^2)-g_1(z,Q^2)\right]
\end{equation}
or finally\footnote{Again, for clarity, we use only the nonsinglet and LO
notation}
\begin{equation}\label{eq.3.22}
\frac{dg_2^{WW}(x,Q^2)}{d\ln Q^2} = \frac{\alpha_s(Q^2)}{2\pi}
\int\limits_x^1 \frac{dz}{z}\,P\left(\frac{x}{z}\right)\, g_2^{WW}(z,Q^2).
\end{equation}
The above formula shows that the twist-2 component of the function $g_2$
obeys the standard DGLAP evolution with the same evolution kernel
as $g_1$. Eq.~(\ref{eq.3.22}) provides predictions for the leading twist
contribution to the $g_2$ and hence enables determination of the higher
twist terms from experimental data as
\begin{equation}\label{eq.3.23}
g_2^{HT}(x,Q^2) = g_2^{EXP}(x,Q^2) - g_2^{WW}(x,Q^2).
\end{equation}
The obtained evolution equation Eq.~(\ref{eq.3.22}) can be
a useful additional tool for the study of the nucleon structure function $g_2$.

\begin{figure}[h!]
\begin{center}
\vspace{-100mm}\includegraphics[width=0.9\textwidth]{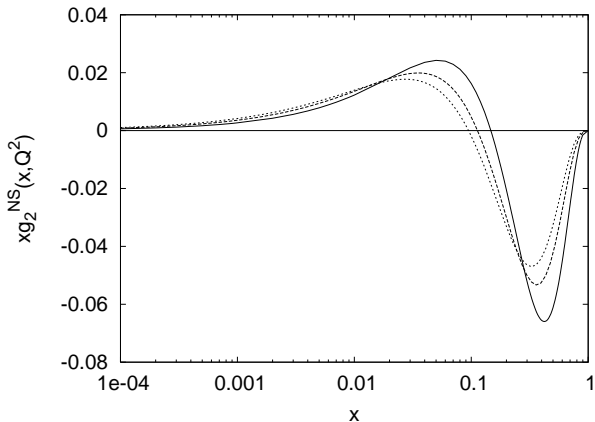}
\parbox{120mm}{\caption
{
The nonsinglet LO contributions to the polarized structure function
$xg_2^{NS}(x,Q^2)$ as a function of $x$ for different $Q^2$: $\rm{1\, GeV^2}$
(solid), $\rm{10\, GeV^2}$ (dashed) and $\rm{100\, GeV^2}$ (dotted).
Parametrization of $g_1$ is given by Eq.~(\ref{eq.3.24}) with
$\alpha = -0.4$.
}}
\end{center}
\end{figure}
\begin{figure}[h!]
\begin{center}
\vspace{-120mm}\includegraphics[width=0.9\textwidth]{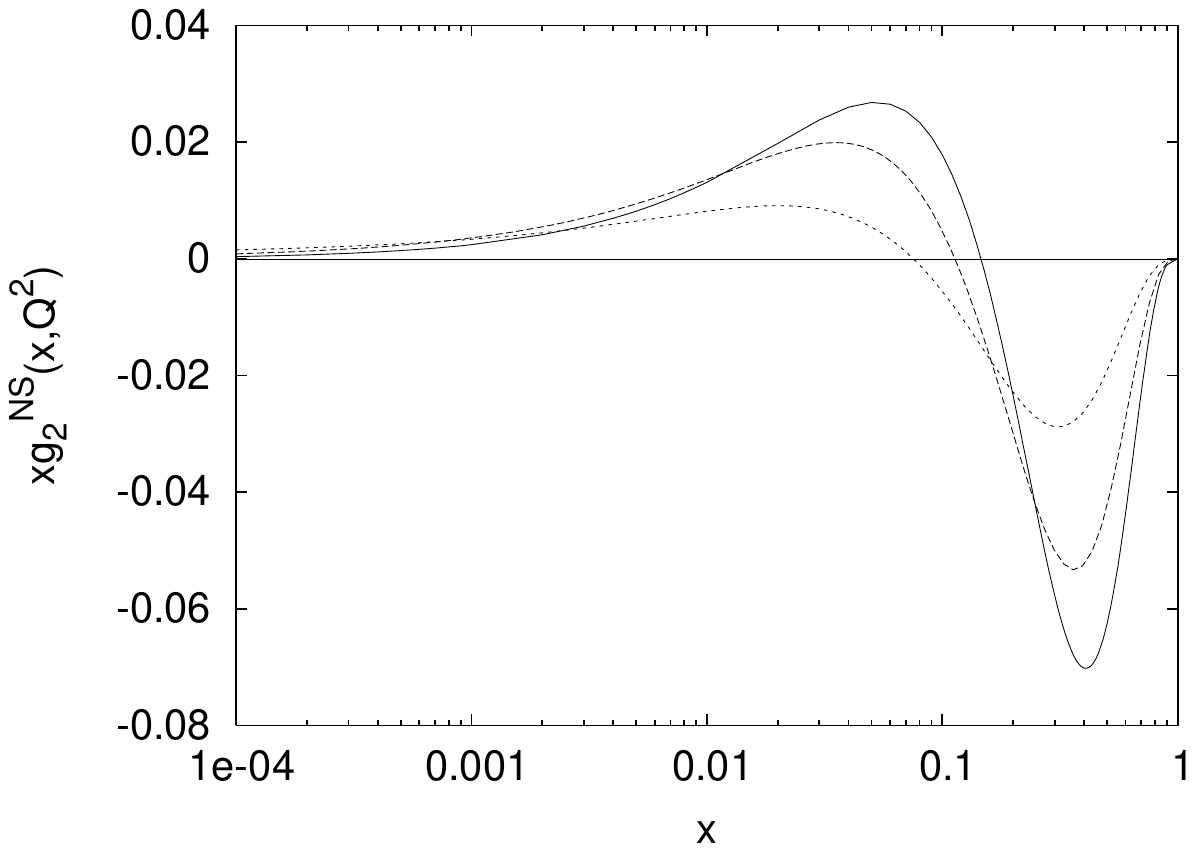}
\parbox{120mm}{\caption{
The nonsinglet LO contributions to the polarized structure function
$xg_2^{NS}(x,Q^2)$ at $Q^2=\rm{10\,GeV^2}$ as a function of $x$ for different
parametrizations of $g_1$, given by Eq.~(\ref{eq.3.24}): $\alpha = 0$ (solid),
$\alpha = -0.4 $ (dashed) and $\alpha = -0.8$ (dotted).
}}
\end{center}
\end{figure}
\begin{figure}[h!]
\begin{center}
\vspace{-100mm}\includegraphics[width=0.9\textwidth]{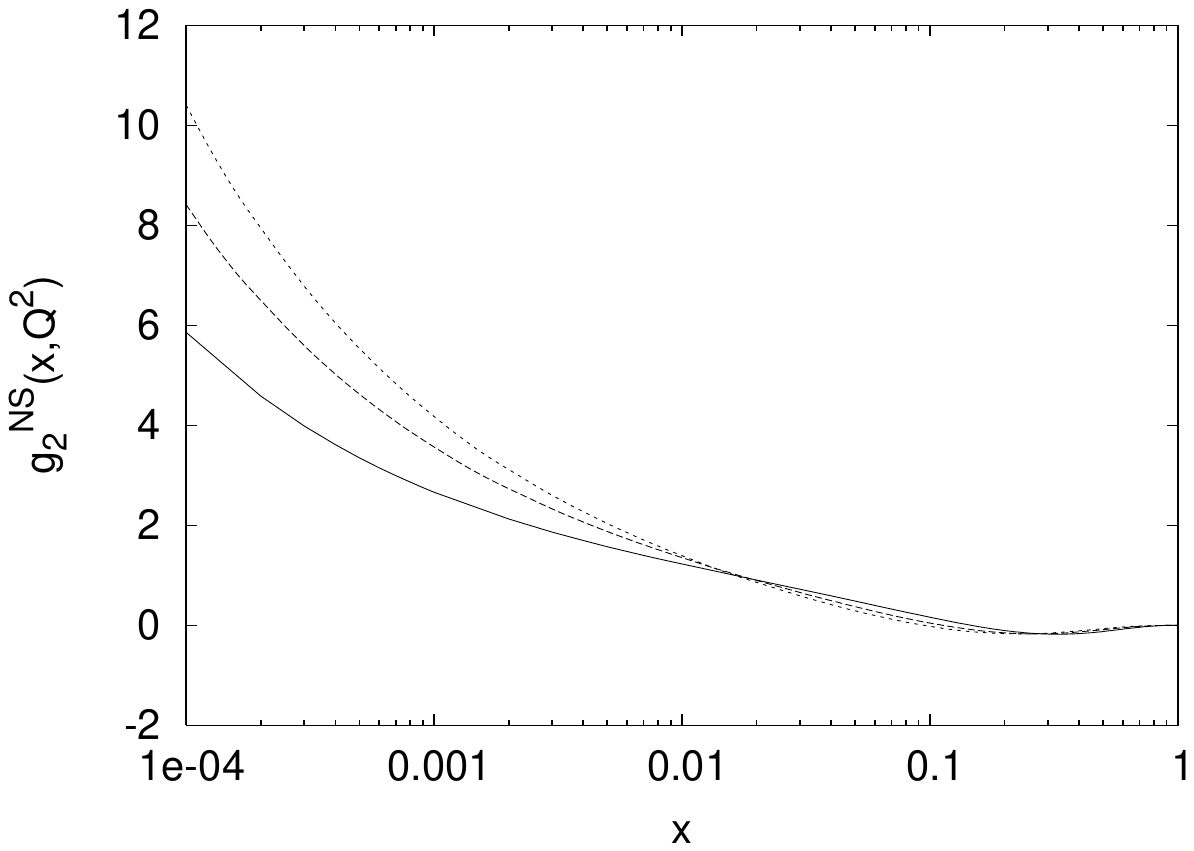}
\parbox{120mm}{\caption{
The nonsinglet LO contributions to the polarized structure function
$g_2^{NS}(x,Q^2)$ as a function of $x$ for different $Q^2$: $\rm{1\, GeV^2}$
(solid), $\rm{10\, GeV^2}$ (dashed) and $\rm{100\, GeV^2}$ (dotted).
Parametrization of $g_1$ is given by Eq.~(\ref{eq.3.24}) with
$\alpha = -0.4$. 
}}
\end{center}
\end{figure}
\begin{figure}[h!]
\begin{center}
\vspace{-120mm}\includegraphics[width=0.9\textwidth]{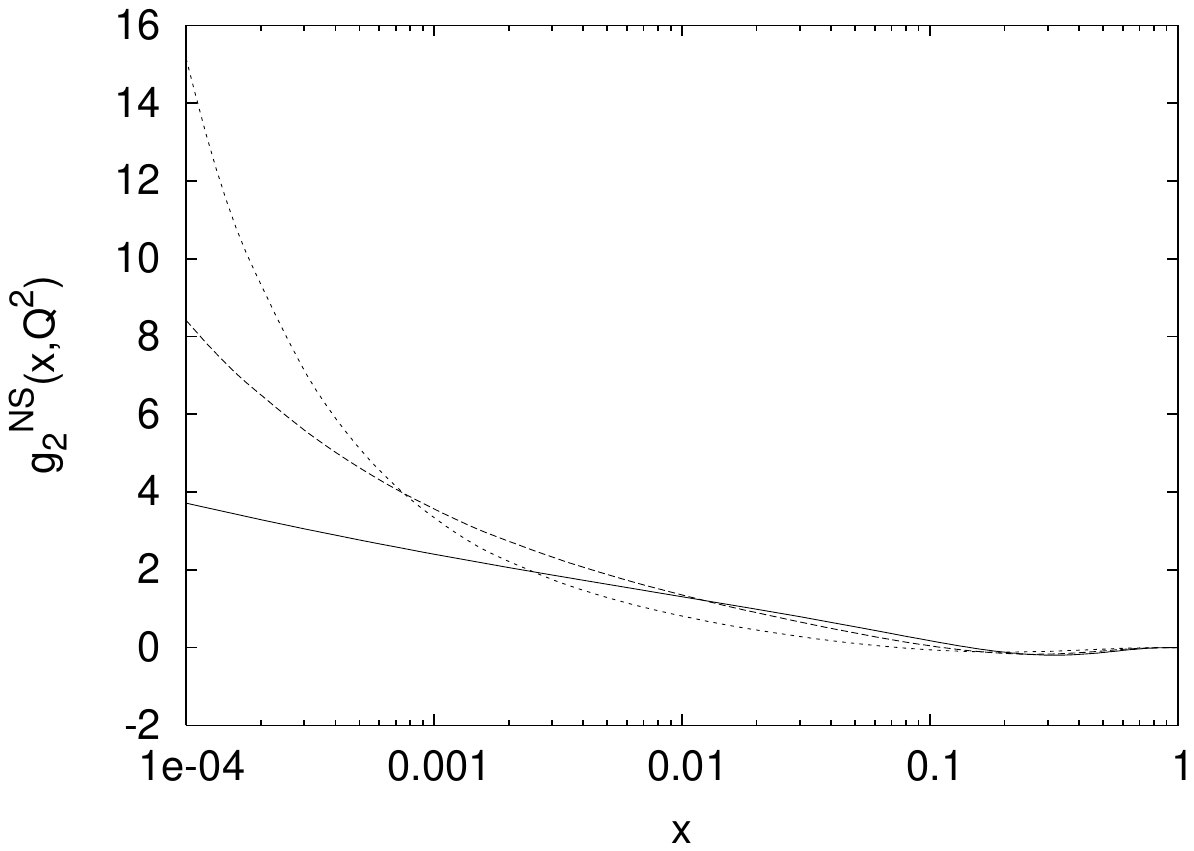}
\parbox{120mm}{\caption{
The nonsinglet LO contributions to the polarized structure function
$g_2^{NS}(x,Q^2)$ at $Q^2=\rm{10\,GeV^2}$ as a function of $x$ for different
parametrizations of $g_1$, given by Eq.~(\ref{eq.3.24}): $\alpha = 0$ (solid),
$\alpha = -0.4 $ (dashed) and $\alpha = -0.8$ (dotted).
}}
\end{center}
\end{figure}
Here, we present numerical solutions of Eq.~(\ref{eq.3.22}).
In Figs.~1,2 we show the nonsinglet contributions to the polarized
structure function $xg_2^{NS}(x,Q^2)$, calculated in the leading order.
We choose the input parametrization of the structure function
$g_1^{NS}(x,Q_0^2)$ at $Q_0^2 = 1 \rm{GeV}^2$ in a general form
\begin{equation}\label{eq.3.24}
g_1^{NS}(x,Q_0^2) = N\;x^{\alpha}(1-x)^{\beta}(1+\gamma x),
\end{equation}
where $\alpha$, $\beta$ and $\gamma$ control the small-, large- and
medium- $x$ behavior of $g_1$ and hence $g_2$, respectively.
The values of the PDF input parameters $\alpha$, $\beta$, $\gamma$
are usually obtained from a global analysis of data and $N$ must be
determined from sum rules constraints. Here, we use a simple fit with
$\beta = 3$, $\gamma = 20$ and different values of $\alpha$ to study its
influence on the evolution of $g_2$. Fig.~1 shows the predictions as
a function of $x$ for different scales of $Q^2$: 1, 10 and 100 $\rm{GeV^2}$.
In the input parametrization of $g_1$ we assume at the initial scale $Q_0^2$
$\alpha = -0.4$. Note that $xg_2^{NS}$ is positive for low-$x$, at about
$x=0.1-0.2$ changes sign and becomes negative for larger $x$. Similar
results were obtained for light and heavy flavor contributions to $g_2$
by J.~Bluemlein, V.~Ravindran and W.L.~van~Neerven \cite{b24}.
This is, as a matter of fact, evidence of agreement with the BC sum rule.
From Fig.~1 one can see also that with increasing $Q^2$, an $x$-intercept
of $g_2^{NS}$ occurs at smaller values of $x$.
In Fig.~2 we compare the predictions for $g_2^{NS}$ for different small-$x$
behavior of the $g_1$ parametrization Eq.~(\ref{eq.3.24}): $\alpha =
0,\,-0.4,\,-0.8$. We find that more singular small-$x$ behavior of $g_1$
implies smaller value of the $x$-intercept of $g_2$. From Figs.~3,4, where
we plot $g_2^{NS}(x,Q^2)$ vs $x$, one can see that the evolution of the
polarized structure function $g_2$ is for small-$x$ very sensitive to the
value of $Q^2$ and, of course, to the assumed input parametrization. For
$x=10^{-4}$ the results of $g_2^{NS}$ can differ by a factor of over 2 from
$Q^2=1\,\rm{GeV}^2$ to $Q^2=100\,\rm{GeV}^2$ for typical input with
$\alpha=-0.4$. Also the input parametrization itself has a large (dominated)
impact on the evolution of $g_2$. Namely, for $x=10^{-4}$ and
$Q^2=10\,\rm{GeV}^2$ the results of $g_2^{NS}$ can differ by a factor of over
4 from the flat input ($\alpha=0$) to the very steep one ($\alpha=-0.8$).\\
Knowledge of the small-$x$ behavior of the structure functions and their
moments is crucial in our understanding of the nucleon structure. Presented
here uncertainty in the determination of the nonsinglet $g_1$ and $g_2$ at
low-$x$ is additionally enhanced in the case of the singlet contributions.
Namely, the results for the singlet $g_1^S$ and $g_2^S$ are very sensitive
on the polarized gluon densities, which are completely unknown. Therefore,
the comprehensive theoretical analysis of the small-$x$ behavior of structure
functions is of a great importance to compensate the lack of the experimental
data  in this region. In the next section we derive some relations between
truncated and untruncated Mellin moments, which can be helpful for such
analyses in the future.

\section{Useful relations between truncated and untruncated Mellin moments}

The evolution equations for the truncated moments Eq.~(\ref{eq.2.9}) are very similar
to those for the parton densities Eq.~(\ref{eq.2.1}). In both cases one deals with
functions of two variables $x$ and $Q^2$ (with additionally fixed index $n$
for moments), which obey the differentio-integral Volterra-like equations.
The only difference lies in the splitting function, which for moments has the
rescaled form Eq.~(\ref{eq.2.10}). This similarity allows one to solve the
equations for truncated moments with use of standard methods of solving the
DGLAP equations. Analysis of the evolution, performed in moment space
according to Eqs. (\ref{eq.2.4})-(\ref{eq.2.7}), when applying to the truncated
moments, implies dealing with such an exotic structure like `Moment of
Moment'. Let us discuss this in detail and introduce some useful relations
involving untruncated and truncated Mellin moments.

There are in literature several methods for the solution of the
integro-differential DGLAP equations. They are based either on the polynomial
expansion or on the Mellin transformation - for review see, e.g., \cite{b25}.
In our previous studies on the evolution of the truncated moments we used the
Chebyshev polynomial technique \cite{b26}, earlier widely applied by Jan
Kwieci\'nski in many QCD treatments - for details see, e.g., Appendix of
\cite{b27}. Using this method, one obtains the system of linear differential
equations instead of the original integro-differential ones. The Chebyshev
expansion provides a robust method of discretising a continuous problem.

An alternative approach is based on the Mellin transformation and the
moments factorization. Taking the $s$-th moment of the evolution equation
(\ref{eq.2.9}), one obtains
\begin{equation}\label{eq.4.1}
\frac{dM^{s,\, n}(Q^2)}{d\ln Q^2}=\frac{\alpha_s(Q^2)}{2\pi}\;
\gamma^{s+n}(Q^2)\, M^{s,\, n}(Q^2),
\end{equation}
where $M^{s,\, n}$ denotes $s$-th (untruncated) moment of $n-$th truncated
moment of the parton density:
\begin{equation}\label{eq.4.2}
M^{s,\, n}(Q^2) = \int\limits_{0}^{1} dx\, x^{s-1}\, \bar{q}^{n}(x,Q^2).
\end{equation}
Analogically to the solution for the parton distribution - 
Eqs. (\ref{eq.2.6a})-(\ref{eq.2.7}), we can write down immediately solutions
for the truncated moments:
\begin{equation}\label{eq.4.3}
M^{s,\, n}(Q^2) = M^{s,\, n}(Q_0^2)
\left[\frac{\alpha_s(Q_0^2)}{\alpha_s(Q^2)}\right]^{\,b\,\gamma^{s+n}}
\end{equation}
and
\begin{equation}\label{eq.4.4}
\bar{q}^n(x,Q^2)=\frac{1}{2\pi i}\int\limits_{c-i\infty}^{c+i\infty}
ds\, x^{-s}\, M^{s,\, n}(Q^2).
\end{equation}
The quantity $M^{s,\, n}$, which is rather exotic and has no physical meaning,
can be replaced by the usual truncated moment $\bar{q}$.
Indeed, the `Moment of Moment' $M^{s,\, n}$ can be written in terms of the
parton density $q(x,Q^2)$ as
\begin{equation}\label{eq.B.1}
M^{s,\, n} = \int\limits_{0}^1 dx\, x^{s-1}\int\limits_{x}^1 dz\, z^{n-1} q(z).
\end{equation} 
For clarity we drop $Q^2$-dependence of the functions.
After simple manipulation with help of the Heaviside function, we can change
the order of integration in Eq.~(\ref{eq.B.1}). Thus, the right-hand-side of
Eq.~(\ref{eq.B.1}) takes the following form
\begin{equation}\label{eq.B.2}
\int\limits_{0}^1 dx\, x^{s-1}\int\limits_{0}^1 dz\, z^{n-1}
\Theta (z-x)\, q(z) = \int\limits_{0}^1 dz\, z^{n-1} q(z)
\int\limits_{0}^1 dx\, x^{s-1} \Theta (z-x).
\end{equation}
Next, absorbing the $\Theta (z-x)$ into the upper limit of integration gives
\begin{equation}\label{eq.B.4}
M^{s,\, n} =
\int\limits_{0}^1 dz\, z^{n-1} q(z)\int\limits_{0}^z dx\, x^{s-1}.
\end{equation}
Finally, performing integration over $x$, we obtain
\begin{equation}\label{eq.4.5}
M^{s,\, n} =
\frac{1}{s}\int\limits_{0}^1 dz\, z^{s+n-1} q(z) = 
\frac{1}{s}\, \bar{q}^{s+n}.
\end{equation}
Now, replacing the $M^{s,\, n}$ in Eq.~(\ref{eq.4.4}) by Eq.~(\ref{eq.4.5}),
we obtain
\begin{equation}\label{eq.4.6}
\bar{q}^n(x,Q^2)=\frac{1}{2\pi i}\int\limits_{c-i\infty}^{c+i\infty}
ds\,\frac{x^{-s}}{s}\, \bar{q}^{s+n}(Q^2).
\end{equation}  
We also find the invert transformation to Eq.~(\ref{eq.4.6}).
Using Eq.~(\ref{eq.4.5}) for the `Moment of Moment' $M^{s-n,\, n}$:
\begin{equation}\label{eq.4.5a}
M^{s-n,\, n} = \frac{1}{s-n}\, \bar{q}^{s},
\end{equation}
we have
\begin{equation}\label{eq.4.7}
\bar{q}^s(Q^2) = (s-n)\, M^{s-n,\, n}(Q^2) = 
(s-n)\int\limits_{0}^{1} dx\, x^{s-n-1}\,\bar{q}^n(x,Q^2).
\end{equation}

Eqs. (\ref{eq.4.5}), (\ref{eq.4.6}) and (\ref{eq.4.7}) are useful
relations between the truncated and untruncated moments.
Particularly Eq.~(\ref{eq.4.6}) seems to have
a large practical meaning and could be applied when the untruncated
moments are known e.g. from lattice calculations. 

\section{Summary}

This paper is a continuation of our earlier studies on the truncated
moments of the parton densities.
We have derived the evolution equation for the double
truncated moments, which is a generalization of those for the single
truncated and untruncated ones. We have obtained the Wandzura-Wilczek
relation in terms of the truncated moments and found new sum rules involving
the structure functions $g_1$ and $g_2$. We have also derived the DGLAP
evolution equation for the twist-2 part of $g_2$. We have presented
numerical predictions for the evolution of $g_2^{NS}$ at different
values of $Q^2$ and for different small-$x$ behavior of the initial
parametrization. We have also derived relations between the truncated and
untruncated Mellin moments, which can be useful in further studies of
spin physics. The method of the truncated moments enables one direct,
efficient study of the evolution of the moments (and hence sum rules) within
different approximations: LO, NLO, etc., in non-spin as well as in spin
sectors. The adaptation of the evolution equations for the restricted
experimentally $x$-region provides a new, additional tool for analysis of
the nucleon structure functions.

Finally, let us list a few of the valuable future applications of the
presented evolution equation and relations for the truncated moments:
\begin{itemize}
\item
Studying the fundamental properties of nucleon structure, concerning
moments of $F_1$, $F_2$ and $g_1$. These are: the momentum fraction carried
by quarks, quark helicities contributions to the spin of nucleon and,
what is particularly important, estimation of the polarized gluon
contribution $\Delta G$ from COMPASS and RHIC data.
\item
Determination of Higher Twist (HT) effects from the moments of $g_2$,
which will be measured at JLab. This is possible via
the generalization of Wandzura-Wilczek relation for the truncated moments
and test of Burkhardt-Cottingham and Efremov-Leader-Teryaev sum rules.
HT corrections can provide information on the quark-hadron duality.
\item
Predictions for the generalized parton distributions (GPDs). Moments of the
GPDs can be related to the total angular momentum (spin and orbital) carried
by various quark flavors. Measurements of DVCS, sensitive to GPDs, will be
done at JLab. This would be an important step towards a full accounting of
the nucleon spin.
\end{itemize}

Concluding, in light of the recent progress in experimental program,
theoretical developments which improve our knowledge of the nucleon
structure functions and their moments are of great importance.\\
\\
We warmly thank K.Golec-Biernat for valuable remarks and suggestions.

\end{document}